\documentclass[12pt]{article}
\usepackage{graphicx}
\usepackage{mathptmx,mathrsfs}
\usepackage{amsfonts,amsmath,amssymb,amsthm}
\usepackage{indentfirst}
\usepackage{cite}
\usepackage{hyperref}
\usepackage{enumitem}
\usepackage{multicol}

\usepackage{bm,graphics,color}

\numberwithin{equation}{section}

\theoremstyle{definition}

\begin{document}

\title{Radiative Corrections in GSDKP}

\author{R. Bufalo$^{1}$\thanks{rodrigo.bufalo@dfi.ufla.br}~, T.R. Cardoso$^{1}$\thanks{tatiana.cardoso@dfi.ufla.br}~, A.A. Nogueira$^{2}$\thanks{andsogueira@hotmail.com}~, B.M. Pimentel$^{3}$\thanks{
pimentel@ift.unesp.br}\\
\textit{$^{1}${\small Departamento de F\'isica, Universidade Federal de
Lavras,}}\\
\textit{\small Caixa Postal 3037, 37200-000 Lavras, MG, Brazil}\\
\textit{$^{2}${\small Universidade Federal do ABC (UFABC),  Centro de Ci\^{e}ncias Naturais e
Humanas (CCNH)}}\\
\textit{\small Av. dos Estados 5001, Bairro Santa Terezinha CEP 09210-580, Santo Andr\'{e}, SP, Brazil}\\
\textit{{$^{3}${\small Universidade Estadual Paulista (UNESP), Instituto de F\'{i}sica Te\'orica (IFT)}}}\\
\textit{\small Rua Dr. Bento Teobaldo Ferraz 271, Bloco II Barra Funda, CEP
01140-070 S\~ao Paulo, SP, Brazil}\\
}
\maketitle
\date{}

\begin{abstract}
We show explicit the first radiative correction for the vertex and photon-photon 4-point function in Generalized Scalar Duffin-Kemmer-Petiau Quantum Electrodynamcis (GSDKP), utilizing the dimensional regularization method, where the gauge symmetry is manifest. As we shall see one of the consequences of the study is that the DKP algebra ensures the functioning of the Ward-Takahashi-Fradkin (WTF) identities in the first radiative corrections prohibiting certain ultraviolet (UV) divergences. This result leads us to ask whether this connection between DKP algebra, UV divergences, and quantum gauge symmetry (WTF) is a general statment.

\end{abstract}


\begin{multicols}{2}
\section{Transition amplitude}

As we know to study quantum processes we need to write the transition amplitude  or, reciprocally, the generating functional (with sources). To construct the transition amplitude we will use the Faddeev-Senjanovic proceeding where we must first do a short study of constraint analysis. The Lagrangian density describing the GSDKP is defined by \cite{AndTese}
\begin{eqnarray}
&&\mathcal{L=}\frac{i}{2}\bar{\psi}\beta ^{\mu }\left( \partial _{\mu }\psi
\right) -\frac{i}{2}\left( \partial _{\mu }\bar{\psi}\beta ^{\mu }\right)
\psi -m\bar{\psi}\psi+\cr\cr
&&+eA_{\mu }\bar{\psi}\beta ^{\mu }\psi -\frac{1}{4}%
F_{\mu \nu }F^{\mu \nu }+\frac{a^{2}}{2}\partial ^{\mu }F_{\mu \beta
}\partial _{\alpha }F^{\alpha \beta},\cr\cr
&&
\end{eqnarray}
where $F_{\mu \nu }=\partial _{\mu }A_{\nu }-\partial _{\nu }A_{\mu }$ is the usual electromagnetic field-strength tensor and $\beta ^{\mu }$ are the DKP matrices that obey the trilinear algebra $\beta ^{\mu }\beta ^{\nu }\beta ^{\theta }+\beta ^{\theta }\beta ^{\nu
}\beta ^{\mu }=\beta ^{\mu }\eta ^{\nu \theta }+\beta ^{\theta }\eta ^{\nu
\mu }\text{ .}$

The transition amplitude can be written in a covariant form in the non-mixing gauge condition \cite{Andgauge}
\begin{eqnarray}
&&Z=N\int DA_{\mu }D\mathcal{\bar{\psi}}D\mathcal{\psi }\times \cr\cr
&&\exp \{i\int
d^{4}x[\mathcal{\bar{\psi}}\left( i\beta ^{\mu }\nabla _{\mu }-m\right)
\mathcal{\psi }+\cr\cr
&&-\frac{1}{4}F_{\mu \nu }F^{\mu \nu }+\frac{a^{2}}{2}\partial ^{\mu }F_{\mu \beta }\partial _{\alpha }F^{\alpha
\beta }+\cr\cr
&&-\frac{1}{2\xi }\left( \partial ^{\mu }A_{\mu }\right) \left(
1+a^{2}\square \right) \left( \partial ^{\mu }A_{\mu }\right) ]\},\cr\cr
&&
\end{eqnarray}
where $\nabla _{\mu }=\partial_{\mu}-ieA_{\mu}$.
It follows from the above result the infinite chain of equations (Schwinger-Dyson-Fradkin equations) that describes completely our theory (without approximations, due to perturbation theory) and the quantum gauge symmetry of this complet equations manifested in terms of Ward-Takahashi-Fradkin identities \cite{GSDKP}.

\section{Quantum gauge symmetry}

Now let us find some identities, arising from gauge symmetry, to analyse before some quantum process of interest . The derivation of (WTF) identities is formally given in terms of the following identity upon the functional generator
\begin{equation}
\left. \frac{\delta Z\left[ \eta ,\bar{\eta},J_{\mu }\right]}{\delta \alpha (x)} \right| _{\alpha =0}=0
\end{equation}
wherein the infinitesimal gauge transformations are given by
$\Psi \rightarrow \Psi +i\alpha (x)\Psi$, $\bar{\Psi}\rightarrow \bar{\Psi}-i\alpha (x)\bar{\Psi}$ and $A_{\mu }\rightarrow A_{\mu }+\frac{1}{e}\partial _{\mu }\alpha (x)$.  
Then the expression for the 1PI-generating functional $\Gamma \left[ \psi ,\bar{\psi},A_{\mu }\right]$ is
\begin{eqnarray}
&&-i\frac{\square }{e\xi }\left( 1+a^{2}\square \right) \partial^{\mu
}A_{\mu }-\bar{\psi}\frac{\delta \Gamma }{\delta \bar{\psi}}+\frac{\delta
\Gamma }{\delta \psi }\psi+\cr\cr
&&+\frac{1}{e}\partial^{\mu }\frac{\delta
\Gamma }{\delta A^{\mu }}=0.\label{partida}
\end{eqnarray}
This is the equation that will supply all the WFT identities.

The first identity we want to point out (in momentum representation) is
\begin{equation}
\Lambda ^{\mu }\left( p,p,k=0\right) =-\frac{\partial }{\partial p_{\mu }}%
\Sigma ^{-1}\left( p\right) .  \label{eq1}
\end{equation}
where we find out that the vertex is related to the DKP self-energy. The second identity we want to point out (in momentum representation) is
\begin{equation}
p^{\mu }\Gamma _{\mu \nu \alpha \beta }(p,p^{\prime },k,k^{\prime })=0.
\label{eq2}
\end{equation}
This identity applies to the study of photon-photon scattering.

As we will see, (WTF) identities will play an important role in building and studying the amplitudes that describe the physical processes.

\section{Radiative corrections}
Before starting the study on the quantum radiative corrections \cite{Func} it is interesting to gain intuition by doing a quantitative analysis of the types of ultraviolet divergences that could appear in (GSDKP). We will conclude, after a brief power count, that the degree of a divergence in a diagram would be $D\dot{=}4-n-\frac{1}{2}n_{m}$, which n=(number of vertex) and $n_{m}$=(number of external mesons lines). With the previous equation, we can classify the types of divergences that have appeared in theory. When D$<$0, we say that the diagram is superficially convergent. For the first radiative correction to the vertex we have $D=0$ (logarithmic divergence) and for the photon-photon process we have $D=0$ too.

\subsection{Vertex}

Now we are going to calculate the first radiative correction associated with the vertex function.

\includegraphics[width=3cm]{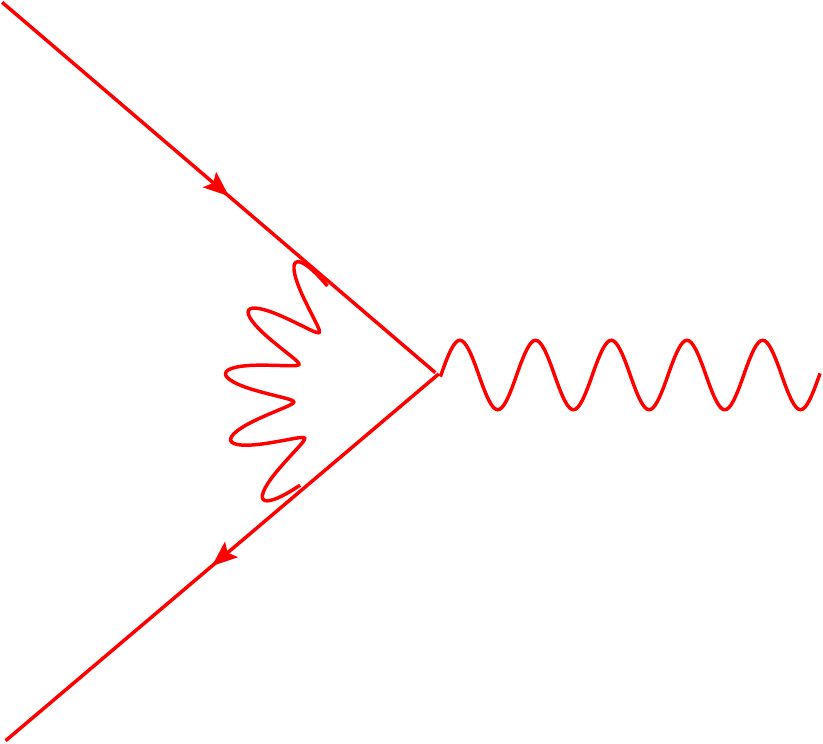}

The vertex part at the
lowest order correction is

\begin{eqnarray}
&&\Lambda ^{\mu }\left( p^{\prime },p\right) =e^{2}\mu ^{4-d}\int \frac{d^{d}k
}{\left( 2\pi \right) ^{d}}\times\cr\cr
&&\beta ^{\sigma }S\left( p^{\prime }-k\right)
\beta ^{\mu }S\left( p-k\right) \beta ^{\nu }D_{\sigma \nu }\left( k\right),\cr\cr
&&
\end{eqnarray}
substituting the expressions for their respective propagators
\begin{eqnarray}
&&\Lambda ^{\mu }\left( p^{\prime },p\right)=\frac{-ie^{2}\mu^{4-d}m_{p}^{2} }{m^{2}}\int \frac{d^{d}k}{(2\pi )^{d}}\times\cr\cr
&&\frac{ \beta ^{\sigma
}[(\hat{p}^{\prime }-\hat{k})(\hat{p}^{\prime }-\hat{k}+m)-(p^{\prime}-k)^{2}-m^{2}]}{[(p^{\prime
}-k)^{2}-m^{2}][(p-k)^{2}-m^{2}]k^{2}\left(k^{2}-m_{p}^{2}\right)}\times\cr\cr
&&\beta ^{\mu }[(\hat{p}-\hat{k})(\hat{p}-\hat{k}+m)-(p-k)^{2}-m^{2}]\beta _{\sigma }.\cr\cr
&&
\end{eqnarray}
This expression may be simplified by making use of the Feynman parametrization, and then be cast into a suitable form
\begin{eqnarray}
&&\Lambda ^{\mu }(p^{\prime },p)=\frac{e^{2}m_{p}^{2}}{(4\pi )^{2}m^{2}}
\int_{0}^{1}dx\int_{0}^{1-x}dy\times\cr\cr
&&\int_{0}^{1-x-y}dz[\frac{A^{\mu}}{b^{4}}-\frac{\eta^{\alpha \nu }B_{\alpha \nu
}^{\mu }}{2b^{2}}-\frac{\Gamma (\frac{\epsilon }{2}) }{4} \left[\frac{4\pi
\mu ^{2}}{b^{2}}\right]^{\frac{\epsilon}{2}}\times\cr\cr
&&\left(\eta^{\alpha \nu }\eta^{\lambda
\theta }+\eta^{\nu \theta }\eta^{\alpha \lambda }+\eta^{\theta \alpha }\eta^{\lambda \nu
}\right) C_{\alpha \nu \lambda \theta}^{\mu }].\cr\cr
&&\label{crver}
\end{eqnarray}
The term $C_{\alpha \nu \lambda \theta }^{\mu }$ presents a
logarithmic divergence. However, by means of using the DKP algebra, one can show that this term actually vanishing because of the identity
$(\eta^{\alpha \nu }\eta^{\lambda \theta }+\eta^{\nu \theta }\eta^{\alpha \lambda
}+\eta^{\theta \alpha }\eta^{\lambda \nu })C_{\alpha \nu \lambda \theta }^{\mu }=0$. Showing therefore that there are no divergences on the vertex part. This result confirms the information contained in one of the (WTF) identity eq. (\ref{eq1}) assuring that the divergence of $\Sigma $ in the term proportional to the mass $m_{p}$  does not spoil the (WTF) identity.

\subsection{Photon-photon}

Finally, let us study the light-light scattering.

\includegraphics[width=5cm]{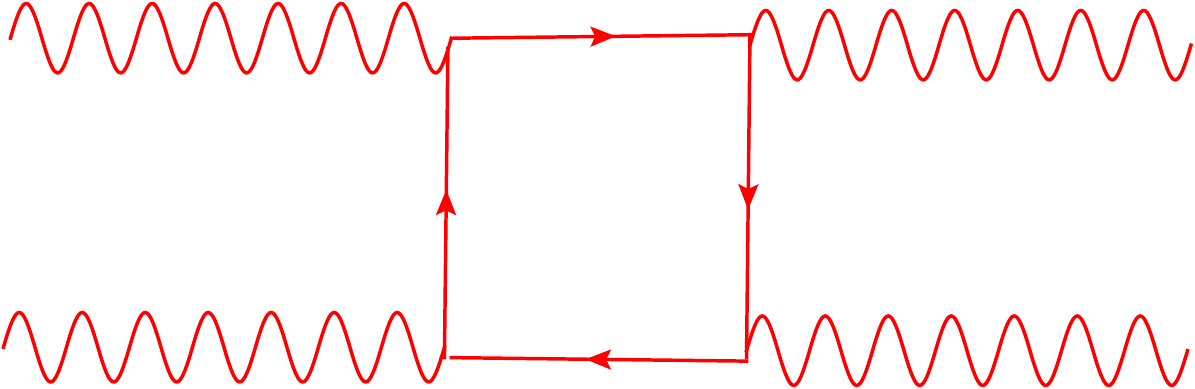}

In view of  Feynman rules we write the amplitude associated with this scattering as follows

\begin{eqnarray}
&&\Gamma ^{_{\mu \nu \lambda \theta }}(p^{\prime },p;p^{\prime },p)=e^{4}\mu
^{4-d}\int \frac{d^{d}k}{(2\pi )^{d}}\times\cr\cr
&&tr\{\beta ^{\mu }S(p^{\prime }+k)\beta
^{\nu }S(k)\beta ^{\lambda }S(p-k)\beta ^{\theta }S(k)\}.\cr\cr
&&
\end{eqnarray}
Thus with the help of Feynman parameters we are led to write explicitly only the term with a possible (UV) divergence

\begin{eqnarray}
&&\Gamma _{UV}^{_{\mu \nu \lambda \theta }}(p^{\prime },p;p^{\prime },p)=\frac{%
ie^{4}}{(4\pi )^{2}m^{4}}tr\{D_{\alpha \sigma \gamma \delta \pi \varpi \rho
\tau }^{\mu \nu \lambda \theta }\}\times\cr\cr
&&(\eta ^{\alpha \sigma }\eta ^{\gamma
\delta }\eta ^{\pi \varpi }\eta ^{\rho \tau }+perm)\times\cr\cr
&&\int_{0}^{1}dx\int_{0}^{1-x}dy(1-x-y)\frac{b\Gamma (-1+\frac{\epsilon }{%
2})}{8[\frac{b^{2}}{4\pi \mu ^{2}}]^{\frac{\epsilon }{2}}},\cr\cr
&&\frac{\Gamma (-1+\frac{\epsilon }{2})}{[\frac{b^{2}}{4\pi \mu ^{2}}]^{\frac{%
\epsilon }{2}}}=[\frac{2}{\epsilon }-\gamma ][1-\frac{\epsilon }{2}\ln (%
\frac{b^{2}}{4\pi \mu ^{2}})].
\end{eqnarray}

In this case the structure representing the light-light scattering must not have (UV) divergences because otherwise it would break the gauge symmetry, seen in eq. (\ref{eq2}). Consequently, as in the first radiative correction of vertex, it is possible to show that DKP algebra prohibits (UV) divergence $tr\{D_{\alpha \sigma \gamma \delta \pi \varpi \rho \tau }^{\mu \nu \lambda
\theta }\}(\eta ^{\alpha \sigma }\eta ^{\gamma \delta }\eta ^{\pi \varpi}\eta ^{\rho \tau }+perm)=0.$

\section{Conclusions}
In this work we show the connection between the DKP algebra, (WTF) identities and the UV divergences for the first radiative correction to the vertex and the photon-photon amplitude. Sometimes, the symmetries of the theory can decrease the  (UV) divergence degree of an amplitude. We conclude that gauge symmetry prohibits ultraviolet divergence for the Feynman amplitudes studied but who assures the previous statement is the DKP algebra. This raises the question of what is role of this algebra in the study of the quantum interaction between the DKP fields (mesons) and Podolsky fields (generalized photons). The next step would be formulate the renormalization program and it is possible to study GSDKP at thermodynamical equilibrium with the Matsubara-Fradkin formalism \cite{Renor,Fintemp}.

\section{Acknowledgement}

R.B. acknowledges partial support from CNPq (Project No. 304241/2016-4) and FAPEMIG (Project No. APQ-01142-17), T.R.C. thanks FAPESP for support, A.A.N. thank PNPD-CAPES for support and B.M.P. thanks CNPq for partial support.

\end{multicols}
\end{document}